\documentclass[twocolumn,trackchanges]{aastex63}

\newcommand{\edensity}{e/{\AA}$^3$}

\newcommand{\change}[1]{#1}         
\usepackage{enumerate}
\usepackage{pbox}

\received{--}
\revised{--}
\accepted{--}

\submitjournal{ApJ}

\shorttitle{Thermal history of forsterite grains from Murchison based on high-resolution tomography}
\shortauthors{Perotti et al.}

\begin{document}

\title{Thermal history of matrix forsterite grains from Murchison based on high-resolution tomography}

\correspondingauthor{Giulia Perotti}
\email{giulia.perotti@nbi.ku.dk}

\author[0000-0002-8545-6175]{Giulia Perotti}
\affiliation{Niels Bohr Institute, Centre for Star and Planet Formation, University of Copenhagen, Øster Voldgade 5--7, 1350 Copenhagen, Denmark}

\author[0000-0002-7004-547X]{Henning O. S{\o}rensen}
\affiliation{Department of Physics, Technical University of Denmark,
Fysikvej, 2800 Kongens Lyngby, Denmark}
\affiliation{Nano-Science Center, Department of Chemistry, University of Copenhagen, 2100 Copenhagen, Denmark}

\author[0000-0002-4618-3178]{Henning Haack}
\affiliation{Maine Mineral and Gem Museum, 99 Main St, Bethel, Maine 04217, USA}

\author[0000-0001-8169-7273]{Anja C. Andersen}
\affiliation{Niels Bohr Institute, Centre for Star and Planet Formation, University of Copenhagen, Øster Voldgade 5--7, 1350 Copenhagen, Denmark}

\author[0000-0003-2825-2886]{Dario Ferreira Sanchez}
\affiliation{Paul Scherrer Institut,
Forschungsstrasse 111, 5232 Villigen PSI, Switzerland}

\author[0000-0003-3333-4421]{Elishevah M. M. E. van Kooten}
\affiliation{Globe Institute, University of Copenhagen,
Øster Voldgade 5--7, 1350 Copenhagen, Denmark}

\author[0000-0002-5885-6317]{Esther H. R. Tsai}
\affiliation{Paul Scherrer Institut,
Forschungsstrasse 111, 5232 Villigen PSI, Switzerland}
\affiliation{Brookhaven National Laboratory, Upton, New York 11973, USA}

\author[0000-0001-6048-3583]{Kim N. Dalby}
\affiliation{Nano-Science Center, Department of Chemistry, University of Copenhagen, 2100 Copenhagen, Denmark}
\affiliation{Haldor Tops{\o}e A/S, Haldor Tops{\o}es All\'e  1, 2800 Kongens Lyngby, Denmark}

\author[0000-0001-8141-0148]{Mirko Holler}
\affiliation{Paul Scherrer Institut,
Forschungsstrasse 111, 5232 Villigen PSI, Switzerland}

\author[0000-0001-9721-7940]{Daniel Grolimund}
\affiliation{Paul Scherrer Institut,
Forschungsstrasse 111, 5232 Villigen PSI, Switzerland}

\author[0000-0002-2184-3360]{Tue Hassenkam} 
\affiliation{Nano-Science Center, Department of Chemistry, University of Copenhagen, 2100 Copenhagen, Denmark}
\affiliation{Globe Institute, University of Copenhagen,
Øster Voldgade 5--7, 1350 Copenhagen, Denmark}

\begin{abstract}
Protoplanetary disks are dust- and gas-rich structures surrounding protostars. Depending on the distance from the protostar, this dust is thermally processed to different degrees and accreted to form bodies of varying chemical compositions. The primordial accretion processes occurring in the early protoplanetary disk such as chondrule formation and metal segregation are not well understood. One way to constrain them is to study the morphology and composition of forsteritic grains from the matrix of carbonaceous chondrites. Here, we present high-resolution ptychographic X-ray nanotomography and multimodal chemical micro-tomography (X-ray diffraction and X-ray fluorescence) to reveal the early history of forsteritic grains extracted from the matrix of the Murchison CM2.5 chondrite. The 3D electron density maps revealed, at unprecedented resolution (64~nm), spherical inclusions containing Fe-Ni, very little silica-rich glass and void caps (i.e., volumes where the electron density is consistent with conditions close to vacuum) trapped in forsterite. The presence of the voids along with the overall composition, petrological textures and shrinkage calculations is consistent with the grains experiencing one or more heating events with peak temperatures close to the melting point of forsterite ($\sim$2100~K), and subsequently cooled and contracted, in agreement with chondrule-forming conditions.
\end{abstract}

\section{Introduction} 
\label{sec:intro}
Carbonaceous chondrites contain materials that provide key insights onto the physical and chemical properties of solar nebula dust grains that later agglomerated into the building blocks of the planets in the Solar System (e.g., \citealt{Scott2007}).
They consist of a fine-grained matrix which harbours spherical chondrules (once molten silicate droplets that formed during transient heating events in the solar nebula) and refractory inclusions (i.e., the first formed solids; \citealt{Connelly2012}). 

The primordial accretion processes in the early protoplanetary disk such as metal segregation, chondrule formation and recycling are still poorly understood.
\change{By following a multimodal approach, we performed a high-resolution pilot study of $\mu$m-sized particles from the Murchison CM2.5 chondrite to gain new insights into their formation processes.
In particular, we used a novel powerful method, Ptychographic X-ray Computed nano-Tomography (PXCT) which, to our knowledge, is here applied to planetary science for the first time.} Apart from adopting PXCT, we employed other non-destructive techniques such as micro X-ray diffraction ($\mu$-XRD) tomography, micro X-ray fluorescence ($\mu$-XRF) tomography, micro X-ray Absorption Near-edge Spectroscopy ($\mu$-XANES) and energy dispersive X-ray spectroscopy (EDXS).

The past twenty years have seen a significant increase in the application of X-ray imaging techniques in geological sciences \citep{Mees2003} and particularly in planetary science \citep{Carlson2006,Uesugi2010} for the characterization of nm- to $\mu$m-sized inclusions incorporated in chondrites. This advancement is mainly due to the capability of X-ray imaging techniques to reconstruct the 3D internal structure of rare chondritic samples without dissecting them (e.g., \citealt{Meier2018,Lo2019}). 

Transmission electron microscopy also provides spatial information and at nanometer resolution, but due to scattering effects it comes with stringent sample requirements, being able to only probe thin samples (100$-$200~nm). Therefore, 3D information is obtained through destructive sectioning. In contrast, the innovative technique used in this study, PXCT, delivers the three-dimensional internal structure of $\sim$10$-$100~$\mu$m thick samples, allowing the majority of samples to be analyzed in their native state \citep{Dierolf2010,Pfeiffer2018}. This is achieved exploiting the higher penetration power of hard X-rays and lensless imaging and it results in a significant improvement in spatial resolution ($\sim$10~nm) compared to X-ray microscopy \citep{Holler2014}. Besides providing qualitative information on the morphology of the specimen, PXCT delivers high-contrast electron density maps to quantitatively study density variations in the sample \citep{Diaz2012}. 

Murchison is one of the most studied of all meteorites and it  represents the most suited sample to benchmark this new multimodal approach. Murchison belongs to the CM (Mighei-type) class of chondrites, which are described as aqueously altered breccias with petrological types ranging from 2.0 (significantly altered) to 3.0 (least altered) \citep{Rubin2007,Kimura2020}. The targeted grains were \change{two} forsteritic grains randomly selected from the Murchison CM2.5 matrix. These grains were not located within or directly adjacent to chondrules, but were placed as isolated grains in the matrix (Fig.~\ref{fig:overview}) and, therefore, provide a unique laboratory to trace the origin of primary dust in carbonaceous chondrites.

The origin of such isolated grains is controversial and has been initially proposed to occur by direct condensation of a gas of solar composition (e.g., \citealt{Fuchs1973,Olsen1978, Weinbruch1993}). However, analogies in the chemical composition of isolated and chondrule grains - such as the presence of silica-rich glass inclusions similar to chondrule mesostasis - led to the interpretation that these isolated grains may form by crystallization within chondrules melts (e.g., \citealt{McSween_1977,Desnoyers_1980,Jones1992}). The latter scenario would imply that isolated forsteritic grains are chondrule fragments, thereby placing new constraints on the genesis of chondrules. More recent literature on this topic (see \citealt{Jacquet2021} and reference therein) is in favour of a genetic link between chondrules and isolated grains. In this context, the high-resolution study carried out in this paper will contribute to ascertain their thermal history.

The article is organized as follows. Section~\ref{sec:methods} describes the suite of methods adopted in this study. Section~\ref{sec:results} presents the results of the structural and compositional analysis. Section~\ref{sec:discussion} discusses the results and potential processes involved in the formation history of the inclusions. Finally, Section~\ref{sec:conclusions} summarizes the main conclusions and lists venues for future applications of X-ray ptychographic tomography to planetary science.

\section{Methods}
\label{sec:methods}

\subsection{Sample preparation}
A small piece of Murchison matrix was gently squashed between two clean glass slides, to release individual grains from the matrix. Suitable forsterite grains with visible inclusions were extracted by an Atomic Force Microscope (MFP3D, oxford instruments) (AFM) tip combined with an optical microscope (Zeiss Axiotech). Using the AFM micro manipulator, the forsterite particles were transferred to a sample holder for the PXCT measurement \citep{Holler2017} and attached using a small droplet of 2-component epoxy (Danalim). 
\vspace{0.5cm} 

\begin{figure*}
    \centering
    \includegraphics[scale=0.8]{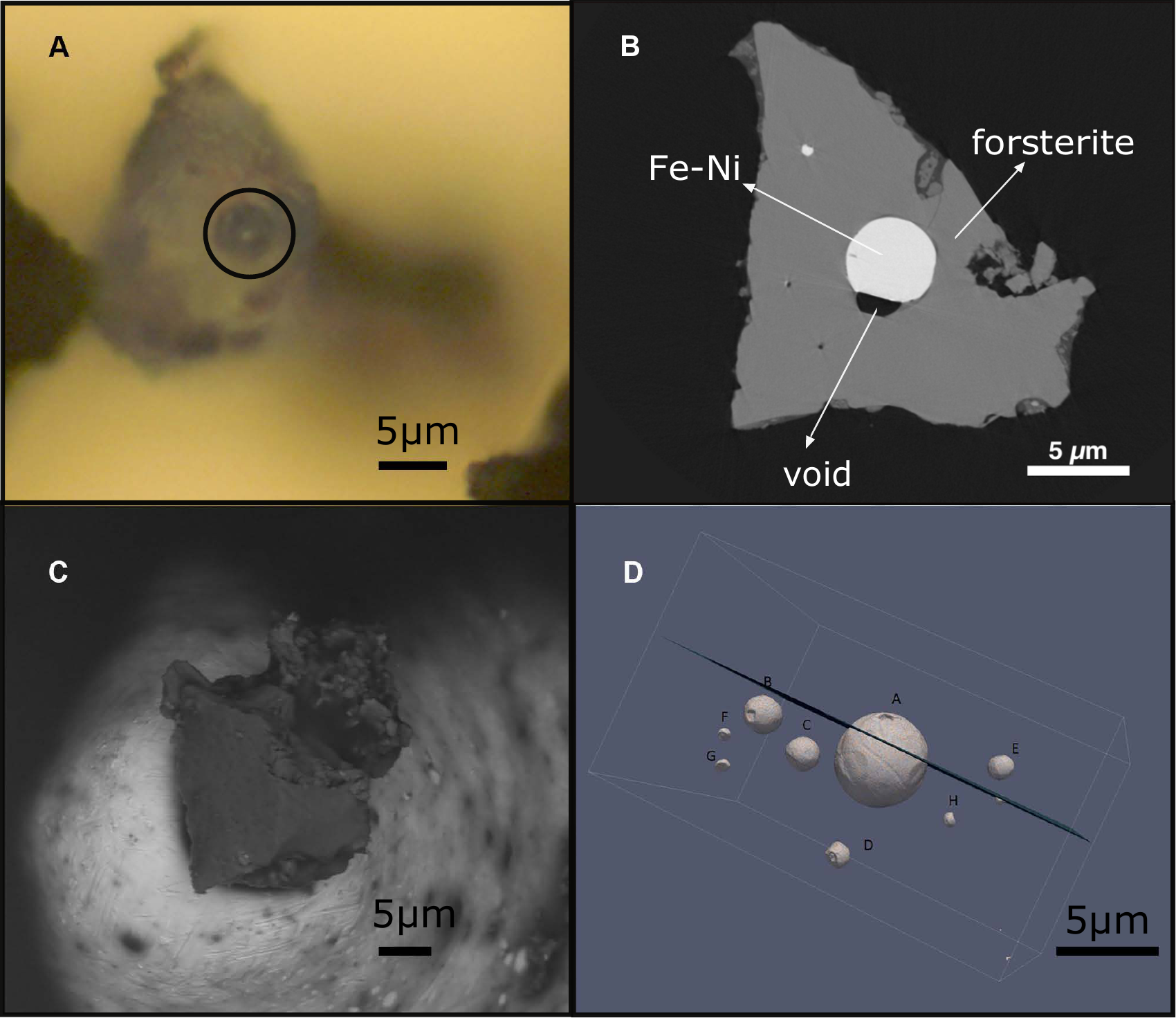}
    \caption{Inclusions from the Murchison chondrite (Sample~\#1). (a) Optical microscopy image of the external morphology of Sample~\#1 showing a transparent reflective metal inside the matrix, consistent with the presence of a larger metal inclusion in the main mineral composing the matrix (empty circle). (b) Ptychographic X-ray computed nano-tomography (PXCT) cross section displaying Fe-Ni inclusions (white) and void caps (black) in a forsterite (Mg$_2$SiO$_4$) matrix (grey). (c) Scanning electron microscope (SEM) image showing the outer structure and the forsterite matrix. (d) 3D PXCT image reconstruction of the internal structure of Sample~\#1, revealing eight spherical inclusions named in alphabetical order from the most (A) to the least voluminous (F).}
    \label{fig:optic}
\end{figure*} 
 
\subsection{Characterisation}

\subsubsection{Scanning electron microscopy and energy dispersive X-ray spectroscopy}
A dual beam FEI Quanta 3D Scanning Electron Microscope (SEM) was used to image samples before tomography and to provide energy dispersive X-ray spectroscopy (EDXS) data. A voltage contrast detector and Oxford X-max 20~mm$^2$ X-ray spectrometer were used to collect the SEM-EDXS data. The samples were not coated, but instead were analysed under low vacuum (60~Pa water vapour) conditions, to minimize charging. The SEM images and EDXS data were obtained using an accelerating voltage of 10~kV and probe current of 8~nA.

\begin{table*}
\parbox{.5\linewidth}{
\begin{center}
\caption{Measured electron densities of the materials identified in the analysed samples with PXCT.}
\label{table:densities}
\renewcommand{\arraystretch}{1.}
\begin{tabular}{llc} 
\hline \hline
Sample     & Location   & Measured value$^a$  \\  
           &            & (\edensity)   \\ \hline \hline
\#1        & host       &     0.97  \\ 
\#1        & inclusion  &     2.04  \\ 
\#1        & aureole    &     0.75  \\
\#1        & void       &    -0.01 \\
\#2        & host       &     0.95  \\
\#2        & inclusion$^b$  & 0.87  \\
\#2        & inclusion$^c$  & 0.84  \\
 \hline 
\end{tabular}
\end{center}
\footnotesize{\textbf{Notes.} $^a$ The estimated uncertainty on the measured values corresponds to 0.02~\edensity. $^b$ The material shown in Fig.~\ref{fig:sample2b} (a). $^c$ The material depicted in Fig.~\ref{fig:sample2b} (b).}
}
\hfill
\parbox{.45\linewidth}{
\begin{center}
\caption{Theoretical electron densities of materials similar to those analyzed in this study.}
\label{table:ref_e_density}
\renewcommand{\arraystretch}{1.}
\begin{tabular}{lc} 
\hline \hline
Reference material & Theoretical value$^a$  \\  
          &           (\edensity)         \\ \hline \hline
Forsterite &  0.97           \\ 
Iron   & 2.18  \\ 
Kaolinite & 0.81 \\
\hline 
\end{tabular}
\end{center}
\footnotesize{\textbf{Notes.} $^a$ The theoretical value represents the electron density calculated for the reference material.}
}
\end{table*}

\subsubsection{Ptychographic X-ray Computed nano-Tomography}
The samples were prepared for a measurement in the flOMNI setup at the cSAXS beamline \citep{Holler2012,Holler2014} of the Swiss Light Source (SLS), Paul Scherrer Institut, Switzerland. They were raster scanned following the Fermat spiral pattern \citep{Huang2014} with 1.2~$\mu$m (except for Sample \#2 where 1.5~$\mu$m was used) step size with a focused X-ray beam with an energy of 6.2~keV and a diameter of approximately 5~$\mu$m. At each point, a coherent diffraction pattern was recorded exposing for 0.1~s using a Pilatus 2M detector~\citep{Henrich2009PILATUS:Applications} that was placed 7.4~m downstream of the sample stage. 
The object pixel size was 21.47~nm, based on the diffraction pattern size (i.e. 400$\times$400 pixels) used for the reconstructions algorithms
~\citep{Thibault2008High-ResolutionMicroscopy, Thibault2012Maximum-likelihoodImaging}. For Sample~\#1 the field of view (FOV) was 36~$\mu$m horizontally and 12~$\mu$m vertically and 700 projections covering 180 degrees sample rotation were taken. An estimated optical 3D resolution (i.e., the scale at which we can resolve objects in the reconstructed image) of 64~nm based on the Fourier Shell Correlation (FSC) \citep{VANHEEL2005250} was achieved. For Sample~\#2 the FOV was 83~$\mu$m horizontally and 14~$\mu$m vertically and a total of 285 projections were employed to reconstruct the 3D image. This lead to a resolution of 86~nm. The quality of the 3D reconstructions for Sample~\#2 was maintained with fewer projections compared to Sample~\#1, spaced by the so-called golden ratio, following the procedure adopted in \citet{Kaestner2011}.

Estimation of the uncertainty of electron density values provided by the PXCT experiments is challenging. From the general variation of the electron densities within well defined volumes, we have estimated the uncertainty to be $\pm$~0.02~\edensity.

\begin{figure*}
    \centering
    \includegraphics[scale=0.95]{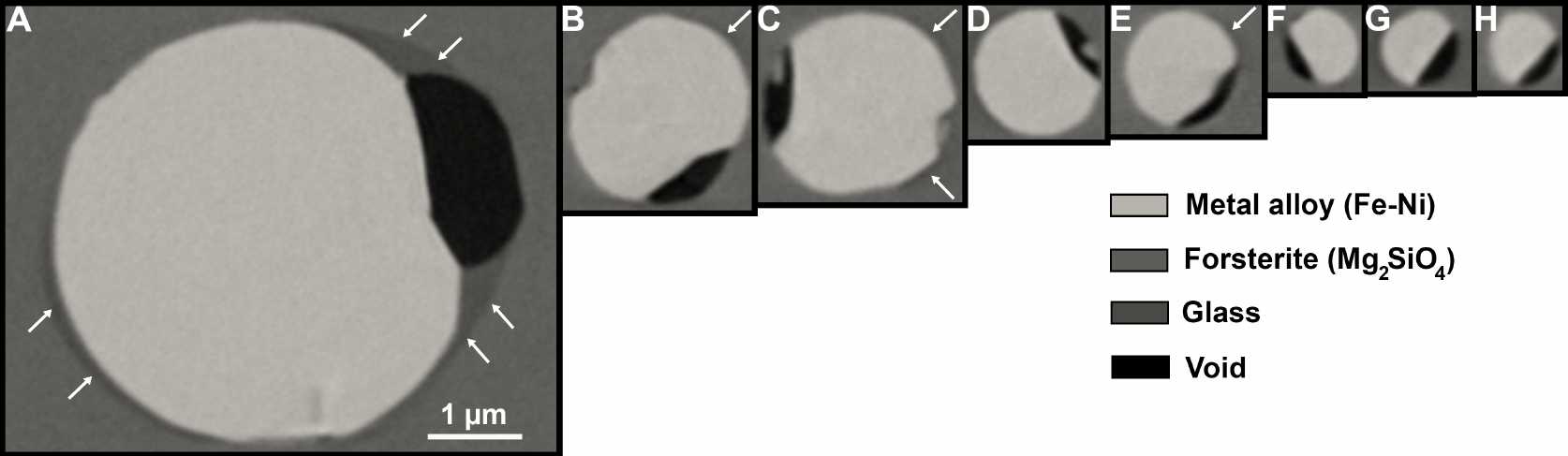}
    \caption{Eight Fe-Ni inclusions and void caps from the Murchison chondrite (Sample~\#1) trapped in a forsterite (Mg$_2$SiO$_4$) matrix. The Fe-Ni inclusions are named in alphabetical order from the most (A) to the least voluminous (H) - see Table~\ref{table:volume}. The void caps are not shown with their correct orientation relative to each other, but to display their maximum extension. The white arrows point to the silica-rich glass.} The images were obtained from PXCT measurements.
    \label{fig:vacuum}
\end{figure*}
\vspace*{\floatsep}
\begin{table*}
\begin{flushleft}
\caption{Volume in $\mu$m$^3$ of the metal particles and of the void caps constituting the eight inclusions (A$-$H) identified in Sample~\#1.}
\label{table:volume}
\begin{tabular}{lcccccccc}
\hline \hline
\multicolumn{9}{c}{Volume ($\mu$m$^3$)} \\ \hline 
 & A    & B    & C   & D   & E & F & G & H  \\ \hline    
Metal      &  48.18 $\pm$ 1       &  3.49  $\pm$ 0.2    &  3.22 $\pm$ 0.2   & 0.94 $\pm$ 0.1  & 0.66 $\pm$  0.1   &  0.27 $\pm$ 0.1   &  0.19  $\pm$ 0.07  & 0.14 $\pm$ 0.05  \\   
Void             &  2.18  $\pm$ 0.15    &  0.13  $\pm$ 0.02   &  0.12 $\pm$ 0.02  & 0.04 $\pm$ 0.01 & 0.03 $\pm$  0.01  &  0.02 $\pm$ 0.01  &  0.04  $\pm$ 0.01  & 0.02 $\pm$ 0.01   \\   \hline 
Ratio            & 0.045  $\pm$ 0.004   &  0.038 $\pm$ 0.01   & 0.037 $\pm$ 0.01  & 0.044$\pm$ 0.02 & 0.052 $\pm$ 0.02  &  0.086 $\pm$ 0.1  &  0.182 $\pm$ 0.2   & 0.120$\pm$ 0.2    \\  \hline 
\end{tabular}
\end{flushleft}
\end{table*}

\subsubsection{Microbeam scanning X-ray fluorescence tomography and X-ray diffraction imaging of Sample~\#1}
The $\mu$-beam scanning X-ray fluorescence ($\mu$-XRF) and $\mu$-beam X-ray diffraction ($\mu$-XRD) tomography data were simultaneously collected at the microXAS beamline (X05LA) of the Swiss Light Source (SLS) at the Paul Scherrer Institut (PSI), Switzerland. Imaging was carried out at 17.2~keV to probe all relevant elements with XRF in the sample, e.g. Fe and Ni and to cover a wide range of the reciprocal space to be able to perform structure determination of the crystalline material in the sample. The energy was calibrated at the K-shell energy of Yttrium using a pure metallic foil of this element. The incident pencil beam was focused to 1~$\mu$m$^2$ using a Kirkpatrick-Baez (KB) mirror system. 
An Eiger4M area detector was used to record the diffraction pattern, and the sample-detector-beam relative position was calibrated using the diffraction pattern of a standard material (LaB$_6$). A home-made python code based on the PyFAI library was used to convert the collected 2D diffraction patterns to 1D diffractograms of intensity vs 2$\theta$ \citep{Ashiotis2015}. Two silicon-drift fluorescence detectors were placed in equal distance on both sides of the sample, in a distance of about 1~cm, perpendicular to the beam direction, in the same polarization-plane as of the incident X-ray photons. The relative incoming flux was recorded using a mini ionization chamber with an Ar atmosphere placed just after the KB mirror system.

The sample was initially scanned in 2D laterally by stepping 0.5~$\mu$m in x and y exposing 200~ms per scanning point. From this 2D projection, we could identify the position of the largest Fe-Ni inclusion. The tomographic scan was scanning across the sample horizontally (90 positions) at the height of the Fe-Ni inclusion stepping  0.5~$\mu$m. At each position we recorded 120 data sets (XRF signal and diffraction images) while rotating the sample over 180$^{\circ}$. In-house developed Python routines based on the ASTRA library using the GPU accelerated SIRT method for parallel beam mode \citep{vanAarle16,Palenstijn2017} were used for the tomographic reconstructions. Structural refinement of the crystalline phases was done by Rietveld refinement (Fig.~\ref{fig:scanning tomo}) against a 1D powder diffractogram that was constructed from $\sim$11000 XRD patterns collected during the scan.

\subsubsection{Microbeam X-ray absorption spectroscopy (XAS)}
Microbeam XAS experiments were performed focusing the X-ray beam on the largest metal inclusion in Sample~\#1. The X-ray energy was calibrated to the Fe K-shell (derived from a pure metallic foil). An X-ray Absorption Near-edge Spectroscopy (XANES) spectrum was recorded, both in transmission and fluorescence mode. 
The best spectrum quality was obtained in transmission mode, which was recorded with a SiC diode. For reference, spectra of an Fe foil, hematite and magnetite were also recorded. 

\section{Results}
\label{sec:results}
\subsection{Morphology of the isolated forsterite grains}
\label{sec:morphology}
\begin{figure*}
    \centering
    \includegraphics[scale=0.8]{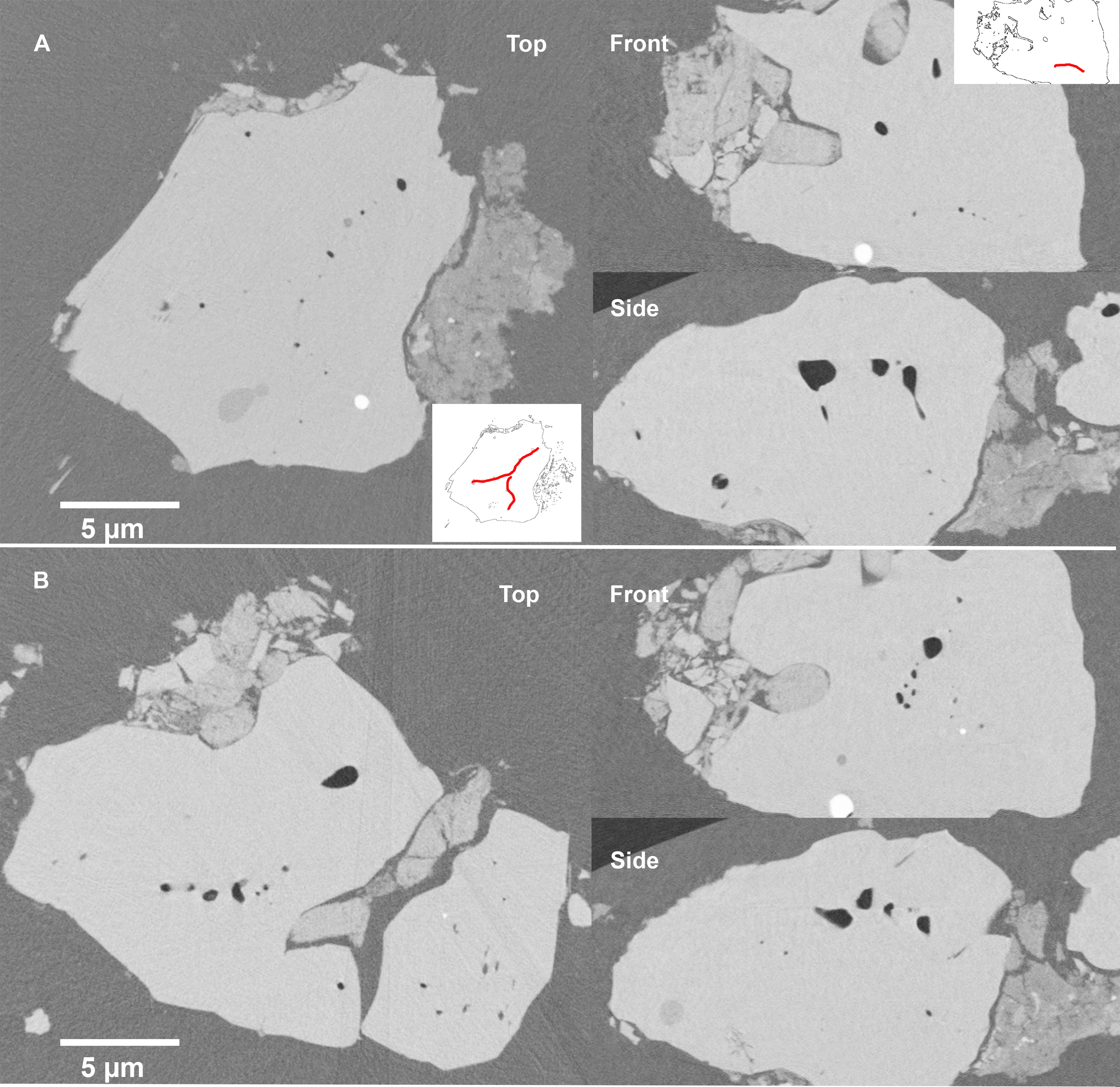}
    \caption{Ptychographic X-ray computed nano-tomography (PXCT) images showing sections of Sample~\#2 from different angles. The white insets in panel (a) depict with solid red lines the voids following a lambda-shape pattern and the inverse parabola, respectively. Panel (b) displays a group of irregular voids.}
    \label{fig:sample2a}
\end{figure*}

\begin{figure*}
    \centering
    \includegraphics[width=4.7in]{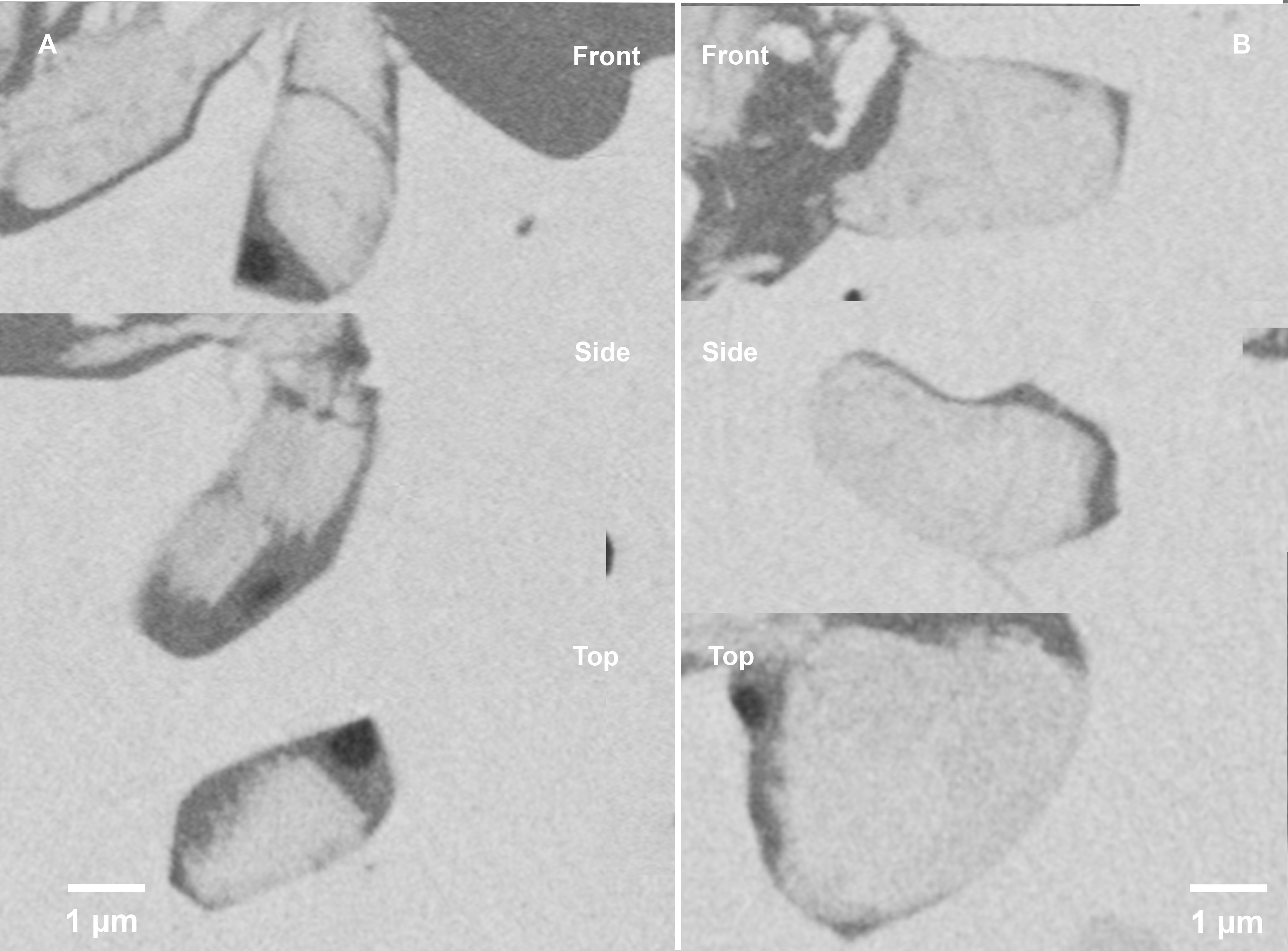}
    \caption{Ptychographic X-ray computed nano-tomography (PXCT) images of partial inclusions of Sample~\#2 from three different orientations. Panels a$-$b show two examples of altered mesostasis within the forsteritic grain.}
    \label{fig:sample2b}
\end{figure*}

Figures~\ref{fig:optic}$-$\ref{fig:sample2b} depict PXCT tomograms of the isolated forsterite grains analysed in this study. PXCT provides 3D information about the electron density distribution, and thus the mass density within the chondritic material, without requiring knowledge of the composition of the samples \textit{a priori} (Sect.~\ref{sec:methods}). The brightest regions in the figures indicate material that has the highest electron density, whereas the darkest regions display the lowest electron density. The two samples show some analogies: (i) each host grain possesses the same electron density ($\sim$0.96~\edensity; Table~\ref{table:densities}) attributable to forsterite (0.97~\edensity; Table~\ref{table:ref_e_density}); (ii) all samples present nm- to $\mu$m-sized inclusions, characterized by higher electron density compared to the host grain. However, the nature of the inclusions is different among the two samples and it will be outlined in the following paragraphs.

Sample~\#1 shows a total of eight spherical inclusions incorporated in the main mineral (Fig.~\ref{fig:optic} a$-$d).
They are characterized by metal domes associated to spherical caps where the electron densities is slightly lower than the ambient air (-0.01~\edensity; Fig.~\ref{fig:vacuum}). We refer to those volumes as ``void caps".

Fig.~\ref{fig:vacuum} presents a slice through the eight spherical inclusions of Sample~\#1, arranged according to decreasing volume. Each metallic spherical dome is accompanied by a void cap that completes the sphere. The volumes of the metal, the void and the metal-to-void ratios are listed in Table~\ref{table:volume}. The uncertainty corresponding to the measured volumes is estimated from the voxel resolution of 21.47~nm.

The diameter of the spherical inclusions varies from a few hundred nanometers up to 5~$\mu$m.  The surfaces of the metal inclusions (A, B and D) have crystalline facets facing the void, whereas they are spherical facing the forsterite. The largest spherical inclusion (A) and other three smaller inclusions (e.g., B, C and E) are accompanied by a minute aureole of silica-rich glass (0.75~\edensity), which occupies a maximum of 1.7~vol\% relative to Fe. 

Fig.~\ref{fig:sample2a} shows a range of PXCT sections of forsterite grains belonging to Sample~\#2. Only a few spherical Fe-Ni inclusions with associated void caps - similar to the ones found in Sample~\#1 (Figs.~\ref{fig:optic}b and \ref{fig:vacuum}) - are present. Sample~\#2 shows a suite of distinct traits. Deformed but rounded voids and spherical inclusions with lower electron densities are trapped in the host grain (Fig.~\ref{fig:sample2a}a,b). In contrast to Sample~\#1, these inclusions are smaller and positioned along lines barely visible in the PXCT images. In fact, these inclusions are not agglomerating towards the center of the grain as in Sample~\#1 (Fig.~\ref{fig:optic}d), but they are aligned along a lambda-shape pattern and an inverse parabola, respectively (Fig.~\ref{fig:sample2a}a). A possible explanation for these inter-grain morphologies is discussed in Sect.~\ref{origin_sample}.   

Fig.~\ref{fig:sample2b} depicts two partial inclusions of Sample~\#2 from three different perspectives. In panels a and b the density of the material inside the open inclusions is slightly less (0.87~\edensity and 0.84~\edensity, respectively; Table~\ref{table:densities}) than the surrounding forsteritic host (0.97~\edensity; Table~\ref{table:densities}). In the open inclusion shown in Fig.~\ref{fig:sample2b}a the material displays crystalline features with stacked platelets. This morphology, together with the electron density, is consistent with phyllosilicate minerals (e.g., clays) forming part of mesostasis aqueously altered to different degrees. In the partial inclusion shown in Fig.~\ref{fig:sample2b}b, which is only 5~$\mu$m from the one shown in Fig.~\ref{fig:sample2b}a, the electron density is similar, which suggests that this material is similar to the material in Fig.~\ref{fig:sample2b}a, but without the stacked platelet features.

\subsection{Elemental composition of Sample~\#1}
\label{sec:elemental composition}
\begin{figure*}
    \centering
    \includegraphics[width=6.5in]{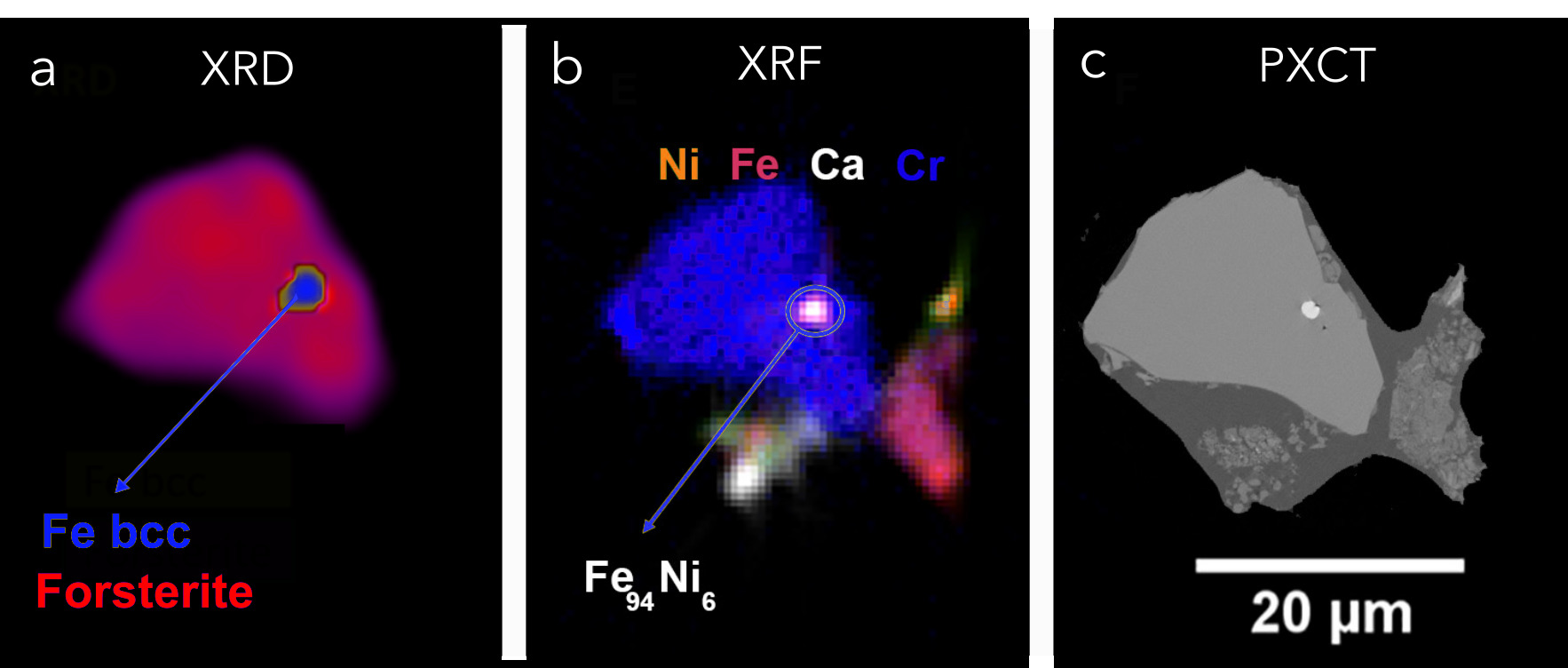}
    \caption{Micro X-ray diffraction ($\mu$-XRD; panel a) tomography, micro X-ray fluorescence ($\mu$-XRF; panel b) tomography and ptychographic X-ray computed nano-tomography (PXCT; panel c) images of Sample~\#1. Supplementary material related to this Figure is shown in Appendix~\ref{appendixA} in Figures~\ref{fig:sample}$-$\ref{fig:scanning tomo}.}
    \label{fig:XRD_XRF_PXCT}
\end{figure*}

The composition and the element distribution of Sample~\#1 (Fig.~\ref{fig:XRD_XRF_PXCT}) were determined using micro X-ray diffraction tomography ($\mu$-XRD), micro X-ray \change{fluorescence} tomography ($\mu$-XRF) and micro X-ray absorption near edge structure ($\mu$-XANES). As the results of these techniques are closely related, they will be presented in a complementary manner.

Both $\mu$-XRD and $\mu$-XRF techniques reveal that the high electron density part of the largest inclusion of Sample~\#1 (i.e., spherical dome in A) is metallic, composed of approximately 94 wt\% Fe and 6 wt\% Ni (Figures~\ref{fig:XRD_XRF_PXCT} and \ref{fig:scanning tomo}). The metal is kamacite (bcc-structure) with the lattice parameter $a$~=~2.86~{\AA} (Fig.~\ref{fig:scanning tomo}). In contrast, the host material surrounding the sphere is a monocrystal of forsterite (Mg$_2$SiO$_4$) with lattice parameters $a$~=~4.75~Å, $b$~=~10.19~Å and $c$~=~5.98~Å. A superposition of all the collected 2D XRD patterns is shown in Fig.~\ref{fig:scanning tomo}~(b), where the data is consistent with monocrystalline structure of the forsterite along with the nanocrystalline nature of the kamacite phase. The spatial distributions of these two crystalline phases were obtained by averaging the normalized reconstructed scattered intensities of the corresponding peaks for the respective phases. Figures~\ref{fig:XRD_XRF_PXCT}b and~\ref{fig:sample} show that the Fe-poor silicate host contains traces of Cr, homogeneously distributed within the crystal, and of Ca, predominantly localized in the matrix surrounding the forsterite grain. The spatial distribution of Ca, Cr, Fe and Ni was obtained through their scattered K$\alpha$ emission lines. 
The SEM-EDXS measurements (Fig.~\ref{fig:EDXS}) confirm that the material hosting the metal inclusions of Sample~\#1 is almost pure forsterite, in agreement with both $\mu$-XANES (Fig.~\ref{fig:XANES}) and $\mu$-XRD data.

Further $\mu$-XANES analyses were performed to assess the potential oxidation of metallic Fe-Ni in the grain, which would indicate aqueous alteration of Sample~\#1 \citep{Rubin2007,vanKooten2018}. The $\mu$-XANES results presented in Fig.~\ref{fig:XANES} do not indicate the presence of Fe oxides inside the forsterite grain and its metal inclusions, suggesting that Sample~\#1 was not subjected to significant aqueous alteration, and hence, represents an unaltered component of the protoplanetary disk. \change{This result is compatible with the forsteritic matrix being particularly resistant to aqueous alteration, and therefore able to largely shield the inclusions within it.}

\section{Discussion}
\label{sec:discussion}

\change{\subsection{On the origin of the observed inclusions}}
\label{origin_sample}
\vspace{0.25cm}
\change{\noindent{Sample~\#1}}\\

The shape of the metal spherical domes of Sample~\#1 and their accompanying void caps (Figs.~\ref{fig:optic}$-$\ref{fig:vacuum}) is consistent with a cooling of the entire sample from above the melting temperature of iron (1811~K; \citealt{Swartzendruber1982}) and \change{around the melting point of} forsterite (2163~K; \citealt{Bowen1935}). This agrees with Sample~\#1 being heated to around 2100~K, at which point the liquid iron would be suspended as spherical droplets in the melt, due to the surface tension between the two liquid materials.
\change{The actual maximum temperature of the melt depends on the purity of the forsterite, as impurities would lower the melting temperature. The SEM-EDXS measurements show no evidence of any elements. The EDXS only investigates composition  within a few hundred nm or micrometers of the surface, but since the tomography of the forsteritic matrix is homogeneous throughout the sample, it is reasonable to assume that the EDXS reflects the composition of the bulk of the particle. If there were other minerals in the host grain, such as fayalite, it would therefore account for less than 1~wt\% according to the detection limit of EDXS. The low content of any fayalite is also inferred from the PXCT data: domains of fayalite in the matrix would have increased the electron density by 33\% with respect to forsterite and aside from the iron spheres we do not observe domains with increased electron density. If we account for 1~wt\% fayalite, the melting temperature of the matrix would decrease by only around 10~K compared to pure forsterite \citep{Ctibor2015}. It is therefore a good first-order approximation to assume the matrix to be predominantly forsteritic with a slightly reduced melting temperature compared to pure forsterite. The peak temperature would therefore likely have been above 2100 K.}

At temperatures where the forsterite would start to solidify, the Fe-Ni alloy would still be liquid and trapped inside the forsterite host as spherical inclusions with a size defined by the equilibrium between the two materials just as forsterite turned solid. As the sample continued to cool, the iron would contract and when the temperature fell to about 1800~K it would also solidify \citep{Swartzendruber1982}. \change{If Fe with 6~vol\% Ni is considered, the melting point of the metallic droplets would be very similar to that of pure Fe (1811~K versus 1790~K; \citealt{Cacciamani2006})}. In this temperature range, the liquid iron would contract more than the surrounding solid forsterite, forming the observed void caps (i.e., shrinkage bubbles; Fig.~\ref{fig:vacuum}), due to the difference in expansion coefficient between the two materials \citep{Watanabe1973,Bouhifd1996}. 
When analysing the void caps of Sample~\#1 (Table~\ref{table:volume}), we assumed that they did not significantly change their sizes during the cooling of the solid iron and forsterite below 1800~K. So that, for instance, the volume of the largest sphere (Fig.~\ref{fig:vacuum}, A) was not subjected to significant variation from the value it had when forsterite solidified and iron was still a liquid droplet. \change{The liquid iron would contract and turn solid causing a combined contraction of about 4.5\% as the droplet cooled from $\sim$2100~K to just below 1800~K \citep{Watanabe1973}}. If we hypothesise that the forsterite contracted as the crystal cooled to just below 1800~K we would have to assume that the volume of the sphere would shrink by about 1.5\% \citep{Bouhifd1996}, so the resulting volume of the void should then be $\sim$3\% of the initial volume.

If our assumption is incorrect, and the volumes of the void caps and solid iron domes changed as the solid cooled even further - in the range from 1800~K to 300~K -  we would have to compensate for the difference in expansion coefficient in this temperature range as well. In this case, forsterite would shrink by about 5\% and iron by about 2\%, with the difference being reversed as forsterite is shrinking more than iron. The iron inclusion should therefore experience high pressure, and if the iron could deform into the void caps, the latter should have been significantly reduced. The fact that the observed void caps are 3$-$4\% of the total volume is therefore consistent with the sample experiencing a peak temperature above the melting point of forsterite and then cooling until both materials were solid just below 1800~K but without subsequent deformation of the iron as temperature was further decreased from 1800 to 300~K. This also agrees with the spherical dome shape of the iron and with the crystalline features facing the accompanying void cap.\\

\change{\noindent{Sample~\#2}}\\

The formation history of the inclusions in Sample~\#2 can be estimated from their morphologies and electron densities revealed by the PXCT data (Fig.~\ref{fig:sample2a}). The spherical inclusions located along lines and the clusters of irregular voids are plausibly the remains of trapped pore space or grain boundaries between individual forsteritic grains (Fig.~\ref{fig:sample2a}). Therefore, these forsterite crystals experienced a sequence of heating events before being incorporated into the CM parent body.

\change{We propose that the forsteritic grains of Sample~\#2 were initially heated in one or more heating events, reaching the melting point of forsterite and subsequently cooled, trapping the pore space into voids, now visible between the grain boundaries, and forming shrinkage bubbles next to the metal. The angular textures suggest that the grains were solid and then fractured into smaller grains during cooling or subsequent impact events of the CM parent body. The voids resisted compaction during these processes. Alternatively, the grains may have formed by the merger of multiple growing olivine grains during crystallization \citep{Jacquet2021}, or have been fractured and partly healed.} 

\change{In addition, the mesostasis observed in Fig.~\ref{fig:sample2b} indicate that the grains of Sample~\#2 interacted with liquid water at some point. These inclusions (originally enstatite or plagioclase) were altered to form clay minerals (e.g., phillosilicates) during secondary alteration on the CM parent body. This scenario would imply that the water was confined to very small volumes surrounding the clay-like minerals as we do not observe similar layered structures in other parts of the samples. Alternatively, it could simply reflect that Sample~\#2 underwent several compaction-impact events, and during one event it was exposed to liquid water.}

This result is consistent with water not homogeneously altering the totality of the CM parent body, but with the action of water confined to a very small volume at distinct spots in the parent body. This agrees with previous studies showing that liquid water was present in the CM parent body, but that the aqueous alteration was heterogeneous \citep{Rubin2007,Trigo2019}.  

\begin{figure*}
    \centering
    \includegraphics[width=7in]{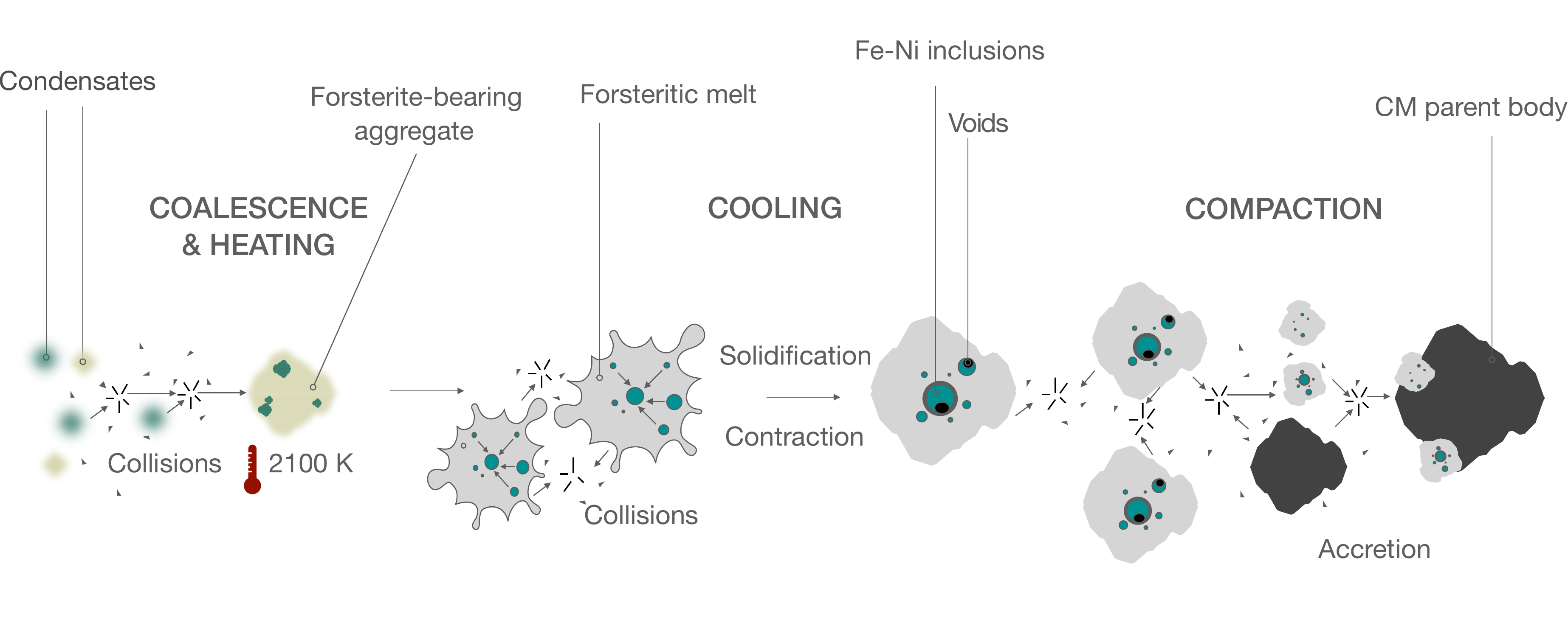}
    \caption{\change{Cartoon displaying a possible formation scenario for the observed inclusions in the isolated forsterite grains of the Murchison CM2.5 chondrite. Heating of the condensates to around 2100~K, melting the forsterite-bearing aggregate. The Fe-Ni particles floated as molten droplets in the melt. Cooling ($10-10000$~Kh$^{-1}$) followed by solidification of the melt and secondly of the Fe-Ni droplets. The contraction of Fe-Ni droplets created shrinkage bubbles (i.e., void caps). Breakup of the aggregate upon cooling via collisional processes, resulting in the fragmentation into smaller grains. Parent body compaction: coalescence of the smaller grains and incorporation into the CM parent body.}}
    \label{fig:cartoon}
\end{figure*}

\subsection{\change{Fayalitic reduction and metal inclusions}}
\label{comparison}
Fe-Ni inclusions resembling the ones observed in Sample~\#1 have previously been found in dusty olivines located in chondrules of ordinary chondrites (Bishunpur LL3.1; \citealt{Rambaldi1981,Leroux2003}, ALH-77015 L3; \citealt{Nagahara1981}, Semarkona LL3.0; \citealt{Jones1997,fu2014,Einsle2016}), as well as in Fe-poor olivine grains isolated in the matrix of CM chondrites (Murchison CM2.5; \citealt{Fuchs1973, Jones1997}, of the Niger(I) C2 chondrite \citep{Desnoyers_1980}, of the ALH-77307 CO3 chondrite \citep{Jones1992}. \change{Compared to the metal inclusions observed in dusty olivines so far, the Fe-Ni inclusions found in Sample~\#1 contain a higher fraction of Ni (up to 6~wt\%). In addition, our sample is generally composed by a lower content of metal ($\leq$1~wt\%), which in turn makes it not ascribable to a dusty olivine.} 

\change{Studies of dusty olivines \citep{Lemelle2000,Leroux2003}} have proposed that metal grains might form via \textit{in situ} reduction by the action of a reducing gas such as hydrogen or by carbonaceous material in chondrule precursors. According to this scenario, metal segregation in chondrites occurs by reduction of fayalitic olivine during chondrule formation \citep{Leroux2003}. This reduction process converts fayalitic olivine into forsterite and it is associated with the formation of: (i) aureoles of silica-rich glass surrounding the Fe-Ni inclusions and (ii) nm-sized silica-rich glass pockets accompanying the metal inclusions.

\change{Although our samples are not dusty olivines, we observe thin aureoles of silica-rich glass around some of the Fe-Ni inclusions in Sample~\#1, plausibly invoking a fayalitic reduction scenario to also explain metal segregation in non-dusty olivine samples (Sect.~\ref{sec:morphology})}. Additionally, we find structures that in shape and size look similar to the aforementioned nm-sized pockets. \change{However, compared to dusty olivines} (e.g., \citealt{Leroux2003,fu2014,Einsle2016}), the pockets reported in Sample~\#1 are not composed by silica-rich glass but are void caps. 

The spherical inclusions A, B, C and E (Fig.~\ref{fig:vacuum}) are surrounded by a sub-micron aureole of silica-rich glass that also envelopes parts of the void caps. \change{If reduction of fayalite is also responsible for the metal segregation in non-dusty olivines there should be double the volume of silica-rich glass compared to that of the metal.} Aside from these thin aureoles we do not register any silica-rich glass in any other parts of Sample~\#1. This implies that if the void caps were filled with silica-rich glass a lot of silica would still be missing from the sample. In the literature, the absence of silica-rich glass has been explained by its efficient volatilization (Si and O move from the bulk to the surface of the grain; \citealt{Lemelle2000}) or by its transport to the outside of the grain \citep{Libourel1995}. If this is the case, it would suggest that our sample was fractured (due to impact) in such a way that the parts containing the silica-rich glass have broken off from the initial grain. 

\subsection{Origin of isolated forsterite grains: condensation vs crystallization?}
\label{origin_crystallization}
The proposed formation scenario for the inclusions in the isolated forsterite grains presented in Sect.~\ref{origin_sample} and elaborated from the analysis of Sample~\#2 is also consistent with the origin of the inclusions in Sample~\#1, although in Sample~\#1 there is no direct evidence of grain boundaries. Rather the spherical shape of the Fe-Ni inclusions and the voids located next to the inclusions, indicate that both forsterite and Fe-Ni alloys experienced one or more heating events reaching temperatures around the forsterite melting point (Fig.~\ref{fig:cartoon}).

The free-floating melt containing Fe-Ni droplets was then cooled and solidified, since the voids do not show a specific orientation (Fig.~\ref{fig:vacuum}). The forsterite likely solidified first, allowing the Fe-Ni alloys to assume the observed spherical shapes. We provide no complex mechanism for how the voids (i.e., shrinkage bubbles) formed other than to propose that they may have originated by the contraction of Fe-Ni droplets upon solidification, consistent with the interpretation previously presented in the literature (e.g., \citealt{Connolly1995}).

\change{The forsterite grains were possibly part of a larger object (e.g., a chondrule) that fragmented and comminute into smaller parts upon collisions during \citep{Jacquet2021} and after cooling} before the ones we designated as Sample~\#1 and \#2 were incorporated into the Murchison chondrite.

Although the forsteritic grains are isolated in the matrix, the overall composition and petrological textures of the grains are commensurate to chondrule-forming conditions \citep{Pack2005}. In addition, the typical micro- and nano-structures found in chondrules reflect the exposure to high temperatures ($\sim$1800$-$2200~K) and cooling rates of 10$-$10 000~Kh$^{-1}$ (e.g., \citealt{Libourel2018}), consistent with the formation conditions proposed for the analysed grains. It is therefore likely that these forsteritic grains crystallized by melting of pre-existing solid material (e.g., a chondrule fragment or processed primary dust) in one or more heating events, similar to a chondrule formation setting \citep{McSween_1977,Jones1992}. 

\change{For the above reasons, a co-formation of chondrules and forsteritic grains prior to the accretion onto the CM parent body is plausible. Alternatively, the isolated grains could also have been derived from heated amoeboid olivine aggregates \citep{Aleon2002,Marrocchi2019}, although less common than chondrules. Due to insufficient data, we are unable to ascertain the exact precursors.} 
In contrast, a condensation origin would imply that the forsteritic melt formed during a gas to solid transition instead of by melting of existing material (e.g., \citealt{Olsen1978,Weinbruch1993}), inconsistent with the presence of shrinkage bubbles and the location of the spherical inclusions in the analysed samples. This conclusion was reached by combining the information from a suite of electron microscopy and X-ray imagining techniques - especially the emerging ptychographic X-ray tomography and high resolution chemical tomography - providing a comprehensive multiple length-scale investigation of the samples. 

\section{Conclusions}
\label{sec:conclusions}
Ptychographic X-ray nano-tomography is employed here for the first time to gain access to the three-dimensional internal structure of \change{two} forsterite grains from the matrix of Murchison at the nanometer scale. The synergy between various X-ray imaging techniques and electron microscopy was exploited to investigate the morphology and elemental composition of the Murchison samples and to get insights into their formation history. The multidimensional data provided quantitative textural and chemical data on the phase assemblage in the Murchison grains. \change{The petrological texture and the shape of the inclusions support a formation scenario in which the grains experienced heating events close to the forsterite melting point, promoting the coalescence of Fe-Ni droplets in a molten matrix and favouring a formation upon crystallization of pre-existing material.}

Additional studies of carbonaceous chondrites at nanometer scale need to be performed, to reconstruct the sequence of events involved in the formation of their parent bodies, as well as to shed light onto the compounds incorporated in them. For instance, such high-resolution studies could support ongoing sample-return missions such as the NASA OSIRIS-REx towards the asteroid (101955) Bennu, a primitive hydrated near-Earth body \citep{Lauretta2015,Lauretta2019a,Lauretta2019b}, and the JAXA Hayabusa-2 towards the asteroids (162173) Ryugu \citep{Morota2020} and (1998) KY26. The present study already shows that plane\-tary science would benefit from ptychographic X-ray tomo\-graphy at the nanoscale, especially for \change{disentangling the origin of chondritic metal whether it is the result of nebular condensation or if it is strictly linked to chondrule formation.}

\section*{Acknowledgements}
The authors wish to acknowledge the anonymous reviewers for the careful reading of the manuscript and the useful comments. We also thank the Swiss Light Source at the Paul Scherrer Institut for providing beamtime at the beamlines cSAXS and microXAS. The Danish Agency for Science, Technology, and Innovation is thanked for funding the instrument center DanScatt. GP acknowledges the financial support from the European Research Council (ERC) under the European Union's Horizon 2020 research and innovation programme (grant agreement No 646908) through ERC Consolidator Grant ``S4F". K. K. Larsen is also acknowledged for kindly providing the Murchison samples. TH thanks the Villum foundation for support under the “Experiment” program Grant number: 17387, and the Danish Council for Independent Research for support under project 1. EHRT was supported by the Swiss National Science Foundation (SNSF) grant number 200021\_152554 and 200020\_169623.

\bibliography{sample63}{}
\bibliographystyle{aasjournal}

\appendix

\section{Supplementary material}
\label{appendixA}

\begin{figure*}[ht!]
    \centering
    \includegraphics[width=6.in]{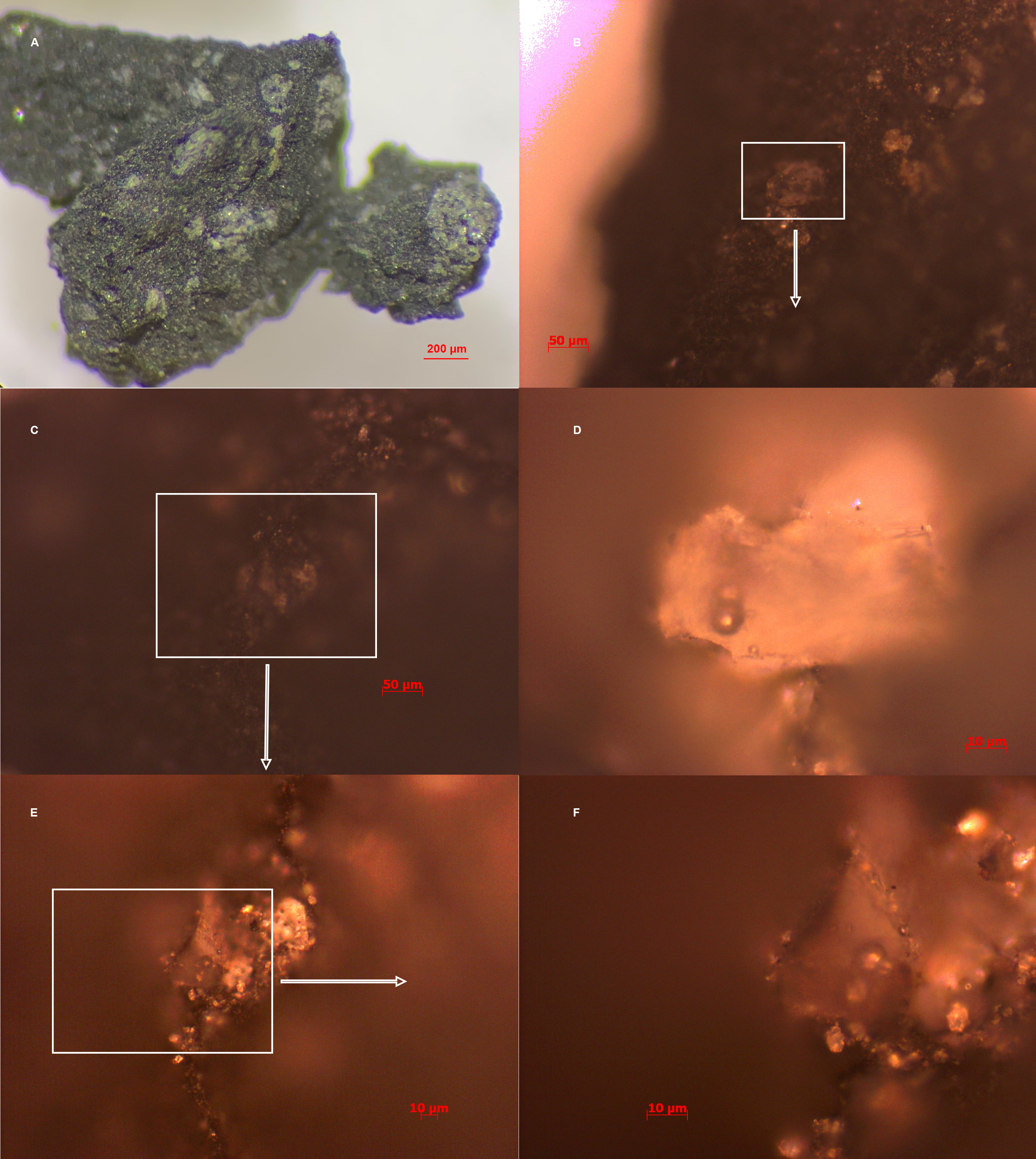}
    \caption{Optical microscopy image of the matrix of the Murchison chondrite from which the samples were extracted.}
    \label{fig:overview}
\end{figure*}

\begin{figure*}[ht!]
    \centering
    \includegraphics[width=6.5in]{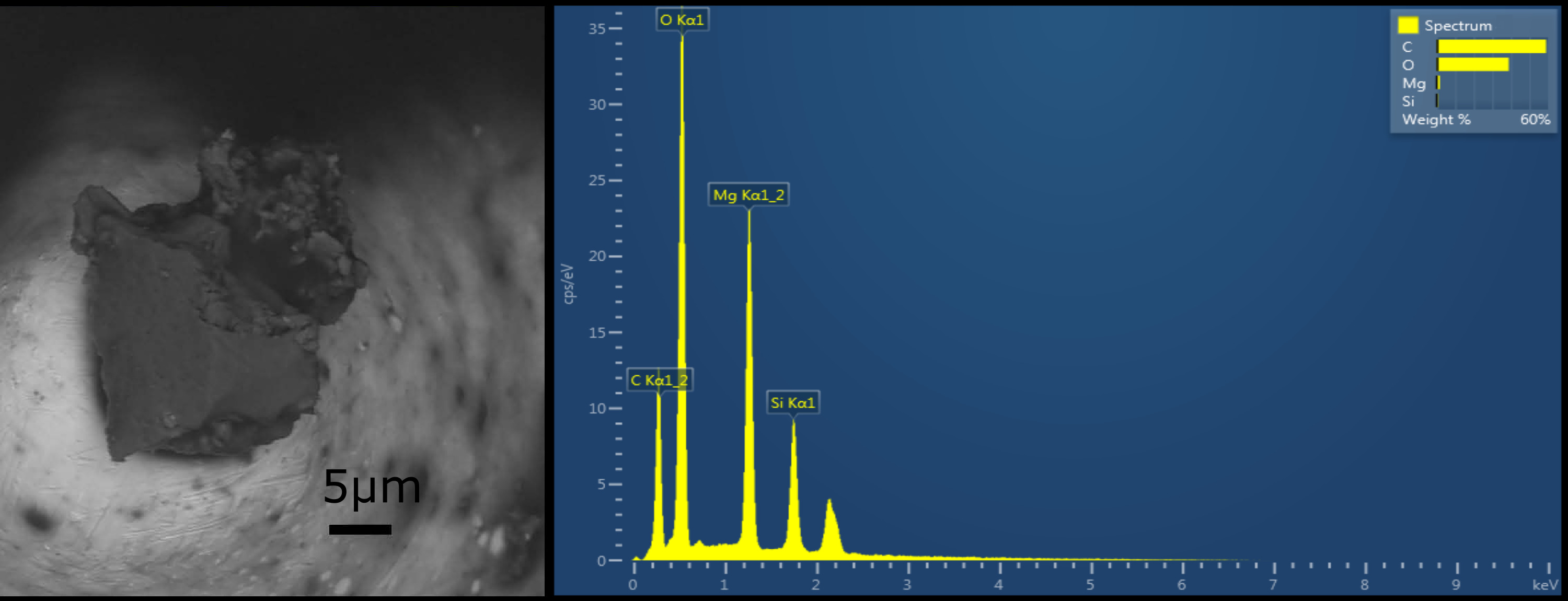}
    \caption{\textit{Left:} Scanning electron microscope (SEM) image of Sample~\#1. \textit{Right:} Energy dispersive X-ray spectroscopy (EDXS) of the matrix of Sample~\#1, confirming the presence of forsterite grains.}
    \label{fig:EDXS}
    \vspace{2cm}
    \includegraphics[width=4in]{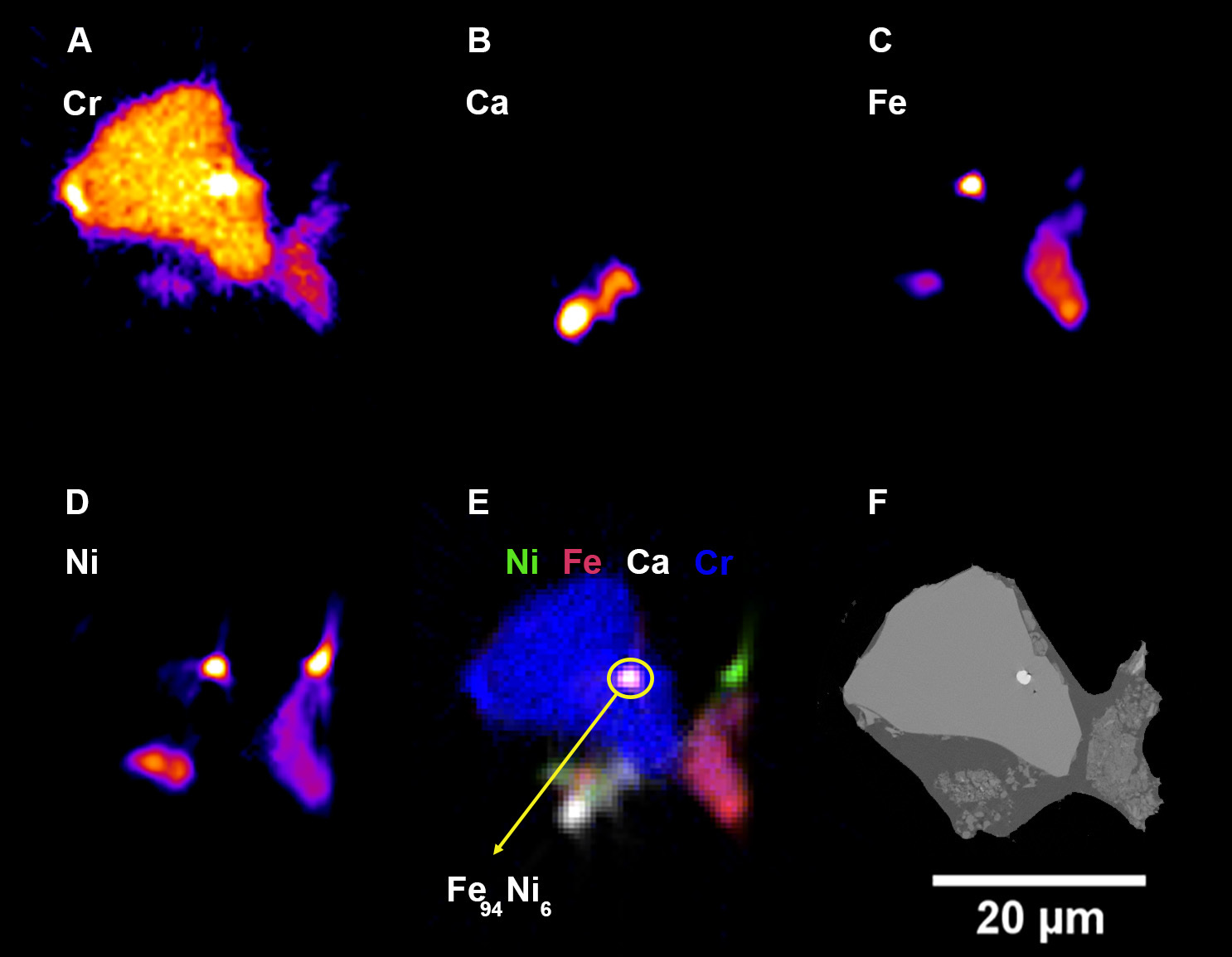}
    \caption{Micro X-ray fluorescence ($\mu$-XRF) tomography image of Sample~\#1. Panels (A)$-$(E) display the distribution of the elements composing Sample~\#1, which is shown as a reference in the PXCT image of panel (F).}
    \label{fig:sample}
\end{figure*}

\vspace{2cm}

\begin{figure}
    \centering
    \includegraphics[width=6in]{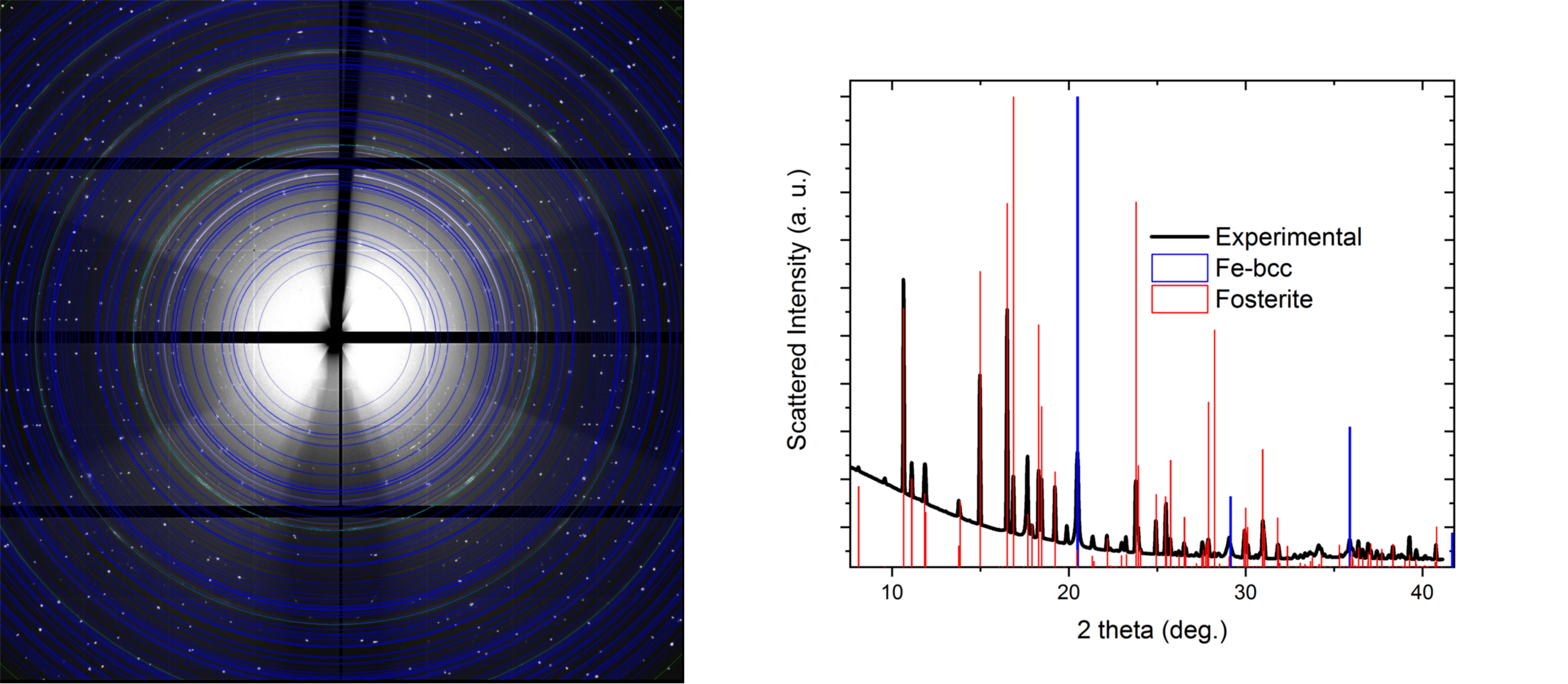}
    \caption{\textit{Left:} Micro X-ray diffraction ($\mu$-XRD) tomography image of Sample~\#1. The colour scheme denotes in blue the forsterite and in green the Fe bcc. \textit{Right:} X-ray diffraction pattern of Sample~\#1.}
    \label{fig:scanning tomo}
\end{figure}

\begin{figure}
\centering
    \includegraphics[width=3.5in]{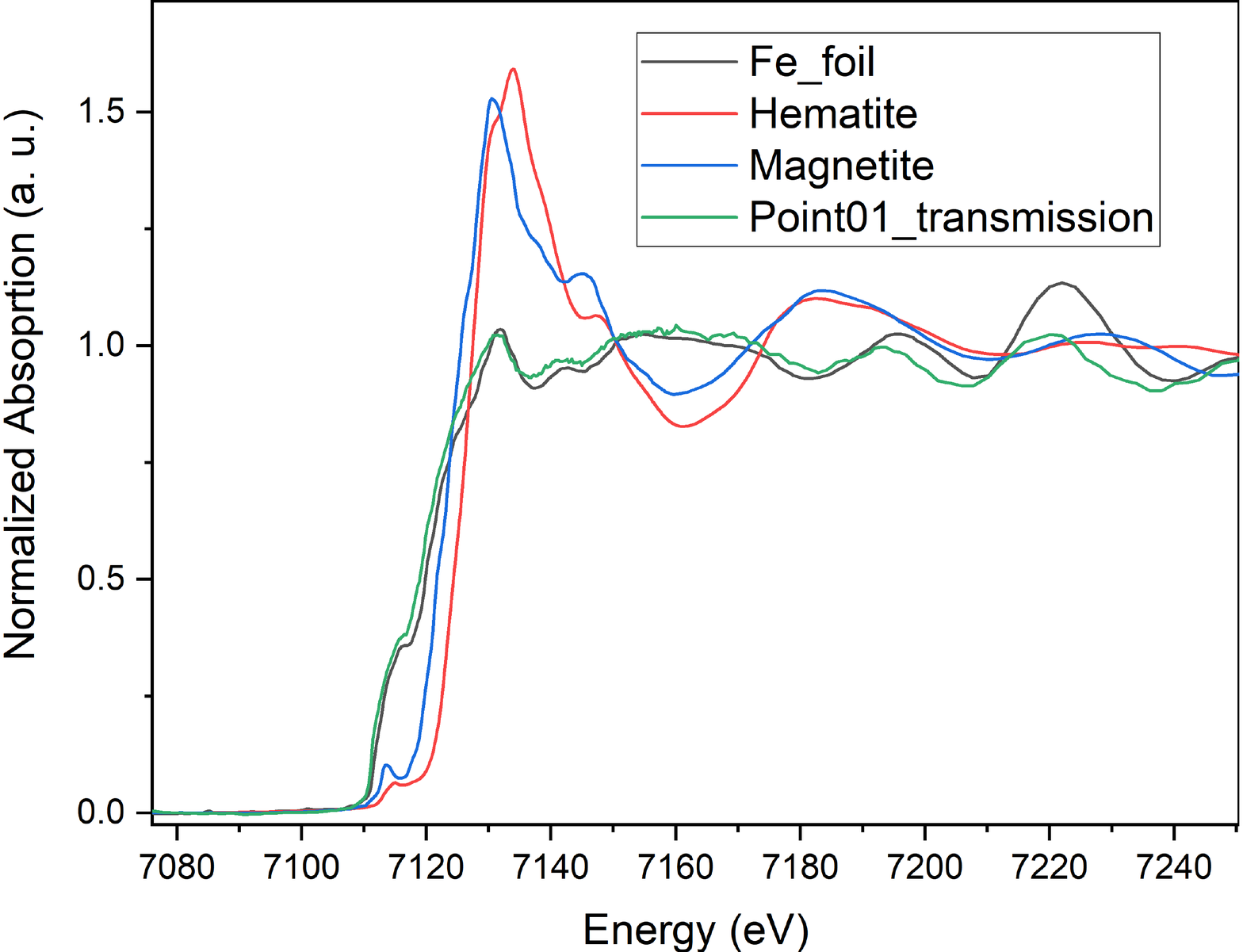}
    \caption{Micro X-ray Absorption Near-edge Spectroscopy ($\mu$-XANES) of Sample~\#1. The Fe foil (grey), hematite (red) and magnetite (blue) line profiles are shown as reference. The point 1 transmission (green) closely follows the profile of Fe foil.}
    \label{fig:XANES}
\end{figure}

\end{document}